\newcommand{\DEt}{\mbox{$\Delta E_{T}$}}

\newcommand{\ETtrue}{\mbox{$E_{\mathrm{T}}^{\mathrm{truth}}$}}
\newcommand{\ETrec}{\mbox{$E_{\mathrm{T}}^{\mathrm{rec}}$}}
\newcommand{\Npart}{\mbox{$N_{\mathrm{part}}$}}

\newcommand{\Rcp}{\mbox{$R_{\rm CP}$}}
\newcommand{\RTwo}{\mbox{$R= 0.2$}}

\newcommand{\RFour}{\mbox{$R = 0.4$}}

\newcommand{\ETtbyf}{\mbox{$E_{\mathrm{T}}^{3\times 4}$}}
\newcommand{\ETsbys}{\mbox{$E_{\mathrm{T}}^{7\times 7}$}}
\newcommand{\ETfcal}{\mbox{$\Sigma E_{\mathrm{T}}^{\mathrm{FCal}}$}}

\documentclass[3p,times]{elsarticle}

\usepackage{ecrc}

\volume{00}

\firstpage{1}

\journalname{Nuclear Physics A}

\runauth{Martin Rybar on behalf of the ATLAS Collaboration}

\jid{nupha}

\usepackage{amssymb}

\usepackage{lineno}

\usepackage[figuresright]{rotating}

\begin{document}

\begin{frontmatter}



\dochead{}

\title{Measurements of jets and jet properties in $\sqrt{s_{NN}} = 2.76$~TeV lead-lead collisions with the ATLAS
detector at the LHC}


\author{Martin Rybar on behalf of the ATLAS Collaboration}

\address{Charles University in Prague, Institute of Particle and Nuclear Physics,
V Holesovickach 2, 180 00 Prague 8, Czech Republic}

\begin{abstract}
Jet quenching in the hot and dense medium created in ultra-relativistic heavy ion collisions is a well-established
experimental phenomenon at RHIC. It has long been anticipated that the LHC heavy ion program would substantially
advance the study of jet quenching by providing access to highly energetic jets and by measuring fully-reconstructed
jets. Immediately following turn-on of the LHC with lead beams, in November 2010, that expectation was fulfilled through the
observation of large di-jet asymmetries that may indicate substantial jet quenching. We will present recent
results from ATLAS aimed to provide further understanding of this phenomenon. Measurements of single jet
production and jet fragmentation in Pb+Pb collisions at $\sqrt{s_{NN}} = 2.76$~TeV will be
presented.
\end{abstract}


\end{frontmatter}


\section{Introduction}
\label{}
\looseness-1 Highly energetic jets produced in a nuclear collision are considered to be a direct probe of hot and dense medium created in the collision. The first measurement of the phenomenon of jet energy loss called "jet quenching"~\cite{quenching} was established at RHIC's experiments. However these observations were limited to the measurement of single, high-$p_{\mathrm{T}}$ particles~\cite{RHIC1}\cite{RHIC2}.

\looseness-1 The first direct observation of the jet quenching using full jet reconstruction was reported in the first ATLAS Pb+Pb paper on dijet correlations~\cite{Asymmetry}. The analysis showed a significant increase of the number of collisions with a large dijet asymmetry with increasing collision centrality and very mild modification of the dijet azimuthal correlation. This energy imbalance strongly suggests the jet quenching. However further measurement was needed to understand the effect. 

\section{Data, jet reconstruction and underlying event fluctuations}
\looseness-1 The data of 47 million minimum bias events corresponding to total integrated luminosity of about 7 $\mu$b$^{-1}$ used in this analysis were recorded with the ATLAS detector~\cite{atlasdetector} during the first LHC Pb+Pb run in 2010 at collision energy of $\sqrt{s_{NN}} = 2.76$~TeV. The event centrality is estimated using the total transverse energy deposited in the forward calorimeters (FCal) that cover $3.2 < |\eta| < 4.9$.

\looseness-1 Jets are reconstructed using the anti-$k_{\mathrm{t}}$ algorithm with different distance parameters $R$ running on calorimeter towers of size $\Delta \eta \times \Delta \phi = 0.1 \times 0.1$. The subtraction of the underlying event (UE) contribution is performed at the finer level of calorimeter cells. Energy to be subtracted is estimated for each longitudinal calorimeter layer and $\eta$ slice separately. The jet energy is corrected for elliptic flow contribution and an additional iterative procedure is used to remove residual effect of the jets on the background estimation. Jet energies are further corrected by an overall multiplicative scale factors obtained from jets simulated with the PYTHIA program embedded in the HIJING event generator. A more detail description of the jet reconstruction can be found in~\cite{Subtraction}.

\begin{figure}
\centerline{
\begin{tabular}{cc}
\includegraphics[width=5.1cm]{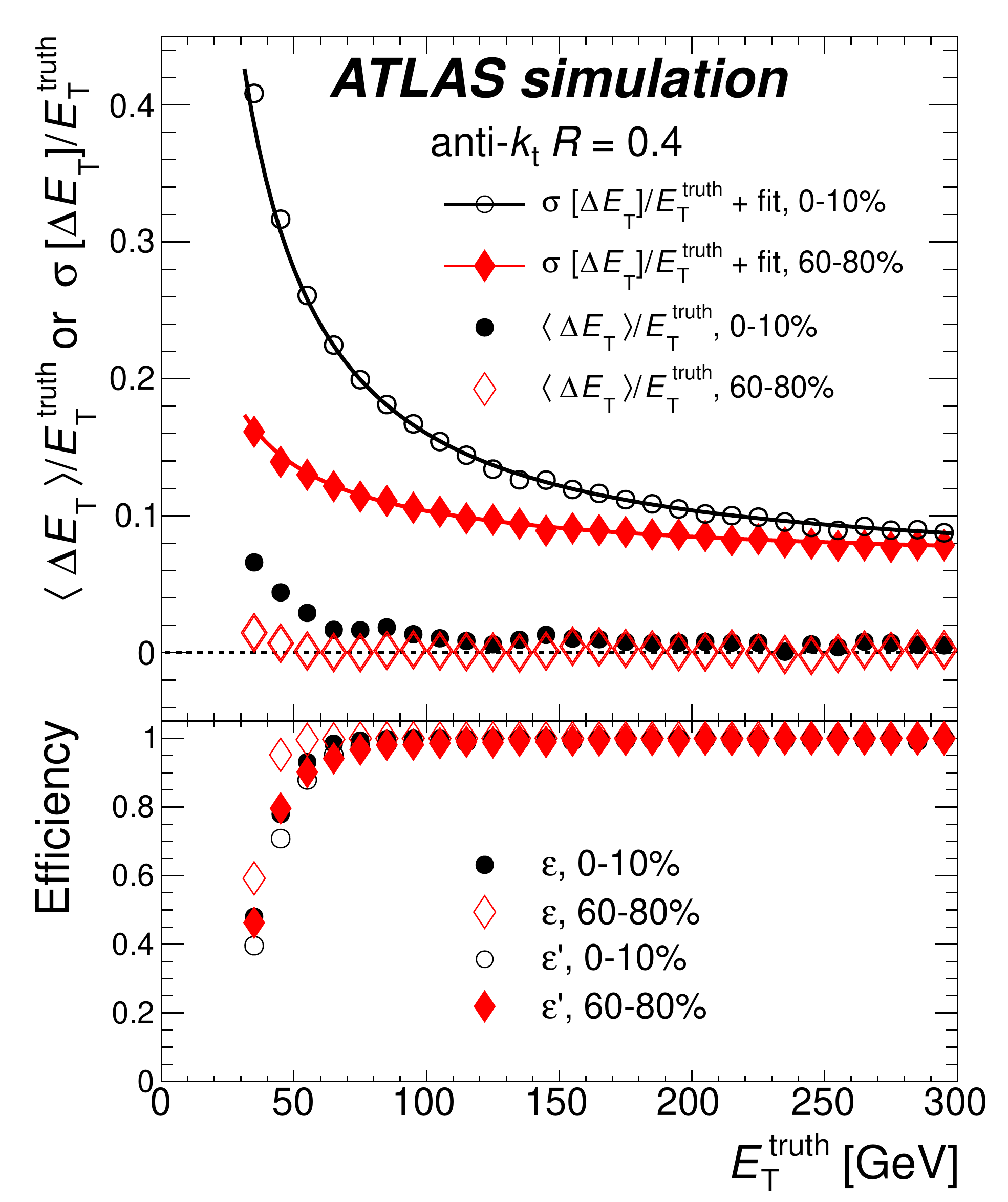} &
\includegraphics[width=8.7cm]{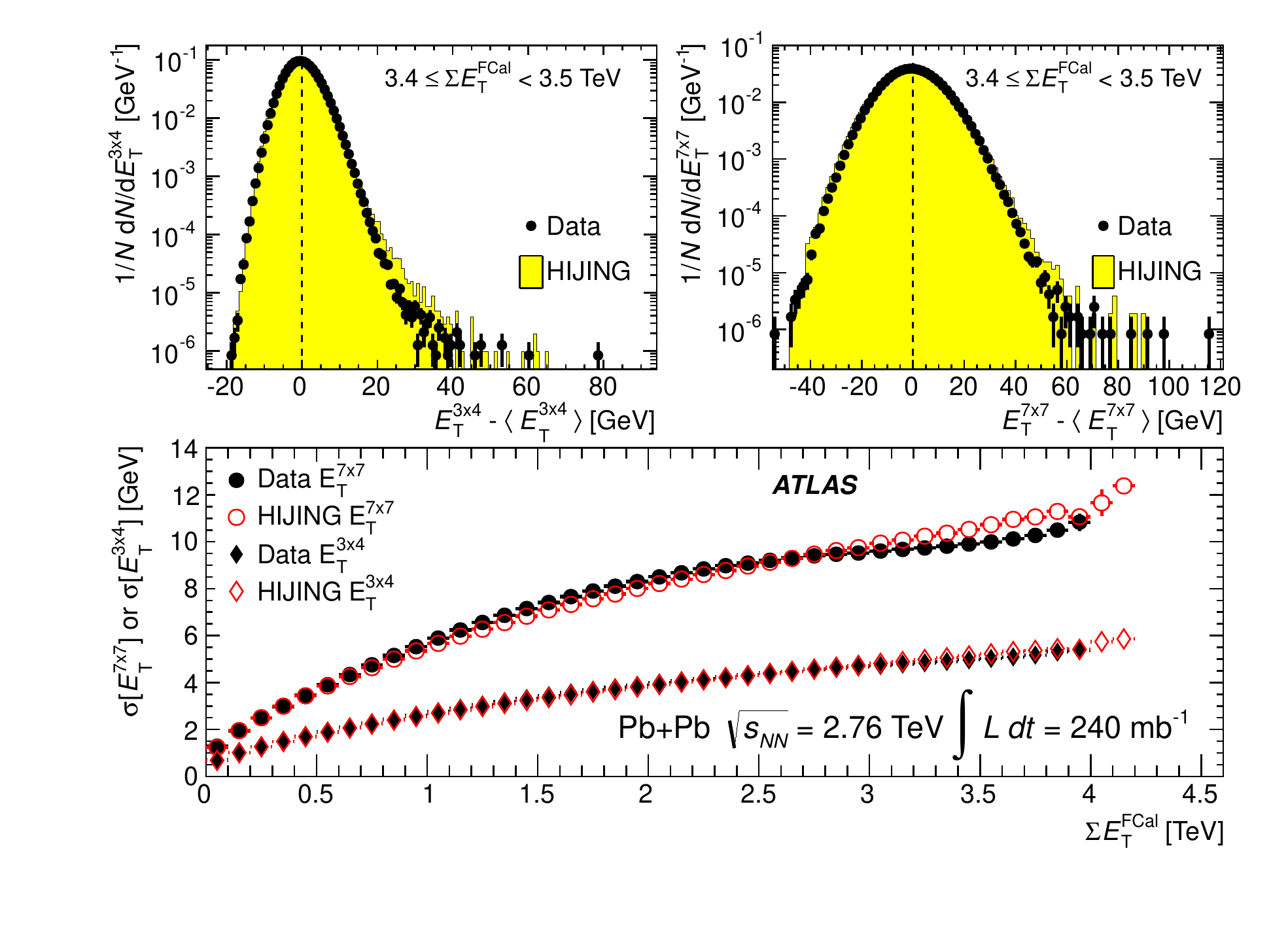} \\
\end{tabular}}
\caption{
Left panels: Jet reconstruction performance in 0-10$\%$ and
  60-80$\%$ centrality bins as a function of truth jet $E_{T}$ for
  $R=0.4$ jets. Top-left: jet energy scale and resolution. Bottom-left: Jet reconstruction efficiencies before ($\varepsilon$) and after ($\varepsilon'$) removal of UE fluctuations. 
  Top-right: Distributions of $\ETtbyf - \langle\ETtbyf\rangle$
  (left) and $\ETsbys - \langle\ETsbys\rangle$ (right)  for data and MC for
  collisions with $3.4$ TeV $\leq \ETfcal < 3.5$~TeV. The vertical lines
  indicate $\ETtbyf - \langle\ETtbyf\rangle = 0$ and $\ETsbys -
  \langle\ETsbys\rangle = 0$, where $\langle\ETtbyf\rangle=26$~GeV and $\langle\ETsbys\rangle=105$~GeV.
  Bottom-right: the standard deviations of the \ETtbyf\ and
  \ETsbys\ distributions, $\sigma[\ETtbyf]$ and $\sigma[\ETsbys]$ in data and HIJING MC as a function of \ETfcal. 
}
\label{Fig1}
\end{figure}

\looseness-1 The left panels of Fig.~\ref{Fig1} summarize the jet reconstruction performance for $R=0.4$ and two different centrality bins. The jet energy resolution (JER) is calculated as $\sigma[\DEt]/\ETtrue$, where $\sigma[\DEt]$ is the standard deviation of the $\DEt \equiv \ETrec - \ETtrue$ distribution. The JER is well described by the formula $a/\sqrt{\ETtrue} \oplus b/\ETtrue \oplus c$, where $a$ and $c$ are the sampling and constant contributions to the resolution. The $b$ term is dominated by UE fluctuations. The jet energy scale (JES) is calculated as the mean energy shift $\langle \DEt\rangle/\ETtrue$. We require $\Delta R<$0.2 matching of calorimeter jets to at least one track jet or a single electro-magnetic cluster with $p_{\mathrm{T}} >$ 7 GeV to identify jets and, thus reject jets from UE fluctuations. Track jets were reconstructed using the anti-$k_{\mathrm{t}}$ algorithm with $R=0.4$ applied to tracks with $p_{\mathrm{T}}>4$~GeV. These requirements restrict our measurement to $|\eta|<$2.1. Efficiency of the jet reconstruction before ($\varepsilon$) and after ($\varepsilon'$) removal of UE fluctuations is presented on the bottom left panel of Fig.~\ref{Fig1}.

\looseness-1 The role of UE fluctuations on the measurement have been extensively studied. Fluctuations are measured as per-event standard deviation of the distribution of energies in groups of towers covering area comparable to area of a jet. The event-averaged fluctuations for $3\times 4$ (comparable to \RTwo\ jets) and $7\times 7$ (comparable to \RFour\ jets) groups of towers as a function of \ETfcal are shown on the bottom-right panel of Fig.~\ref{Fig1}. Two top-right panels of Fig.~\ref{Fig1} show distributions of $E_{\mathrm{T}}$ for $3\times 4$ and $7\times 7$  groups of towers for a narrow range of \ETfcal. A good agreement between data and fully-simulated HIJING events was found. The value of the $b$ parameter obtained from the fluctuations analysis is in a good agreement with the result obtained from the fit of JER. More details can be found in~\cite{RMS}.

\begin{figure}[h]
\centerline{
\includegraphics[width=8.9cm]{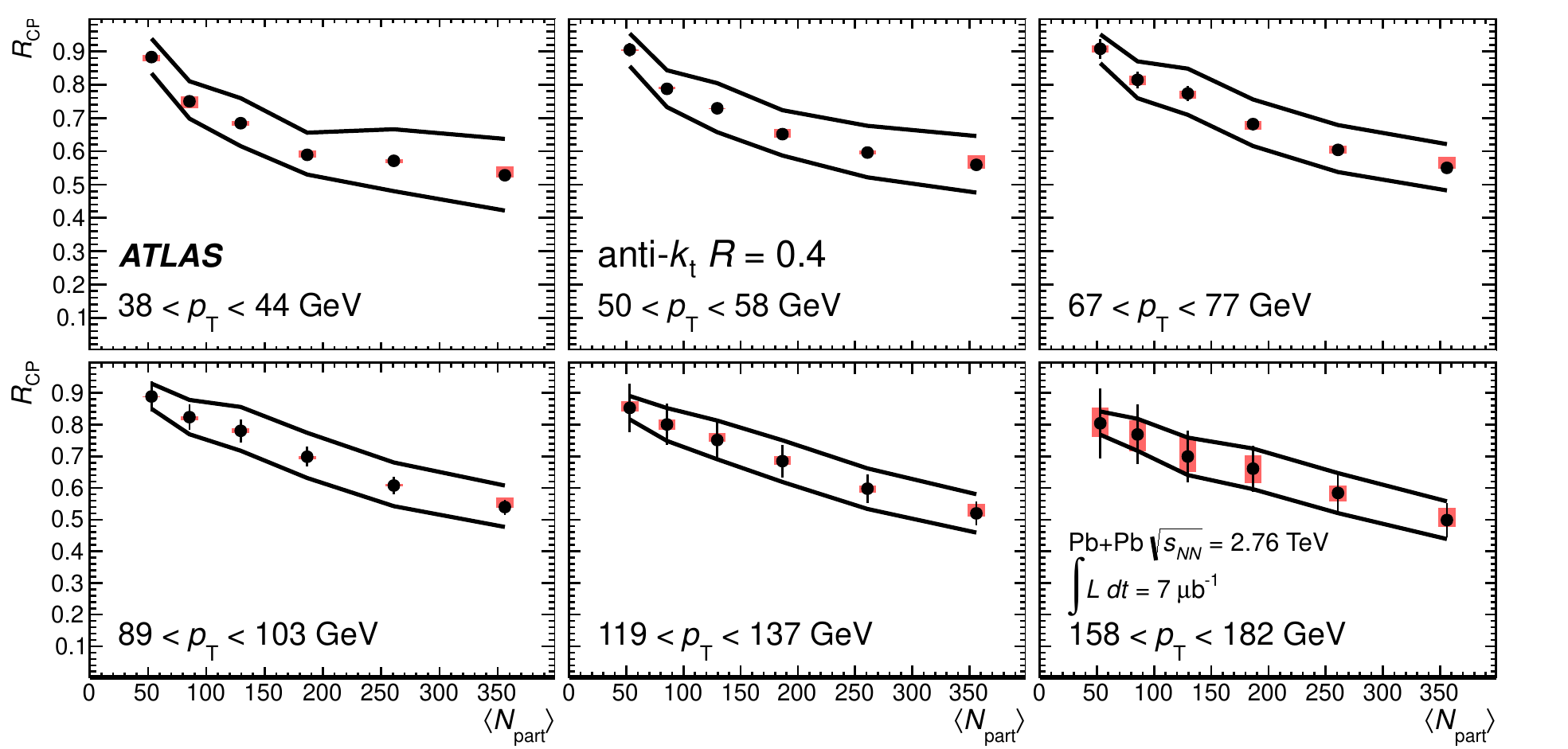}
}
\caption{
\Rcp\ as a function of \Npart\ for \RFour\ jets in six different $p_{\mathrm{T}}$ bins.
The band between the black lines indicates fully correlated uncertainties, shaded boxes partially correlated uncertainties and the horizontal errors indicate systematic uncertainties on \Npart.
}
\label{Fig2}
\end{figure}

\section{Jet suppression}

\looseness-1 We measured the transverse momentum jet distributions in the jet $p_{\mathrm{T}}$ range $38$ GeV $< p_{\mathrm{T}} < 210$ GeV for different values of $R$. Spectra were corrected for increasing jet rate due to collision geometry by scaling with the number of binary collisions $N_{coll}$ obtained from the Glauber model. The spectra were corrected for bin migrations resulting from energy modifications due to detector and UE effects using the SVD unfolding procedure~\cite{SVD}. Jet $R_{\mathrm{CP}}$ was calculated as the ratio of unfolded $p_{\mathrm{T}}$ spectra divided by $N_{coll}$ in a given centrality bin to the same quantity in the peripheral 60-80\% bin.

\begin{figure}[h]
\centerline{
\begin{tabular}{cc}
\includegraphics[width=4.6cm]{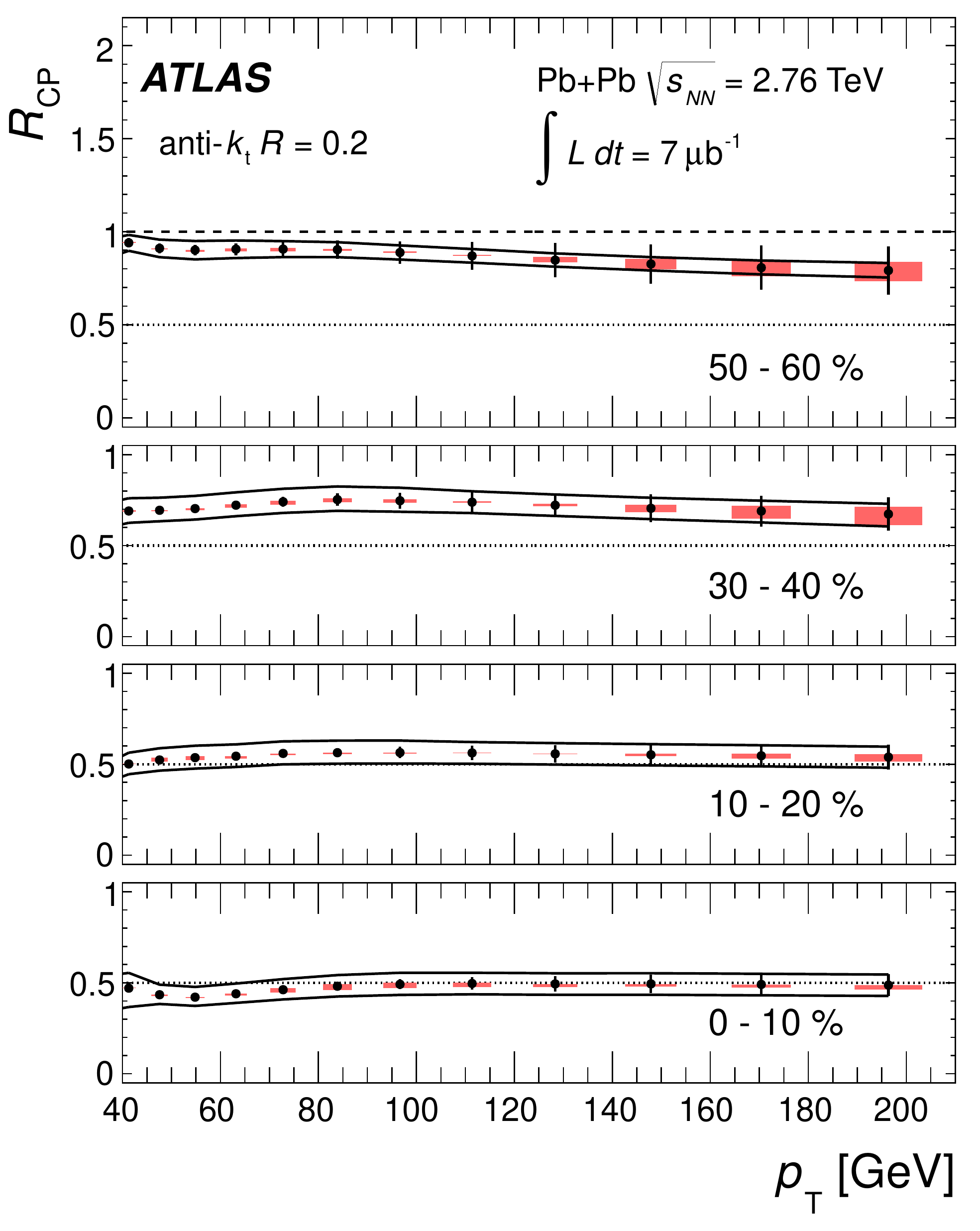}
\includegraphics[width=4.6cm]{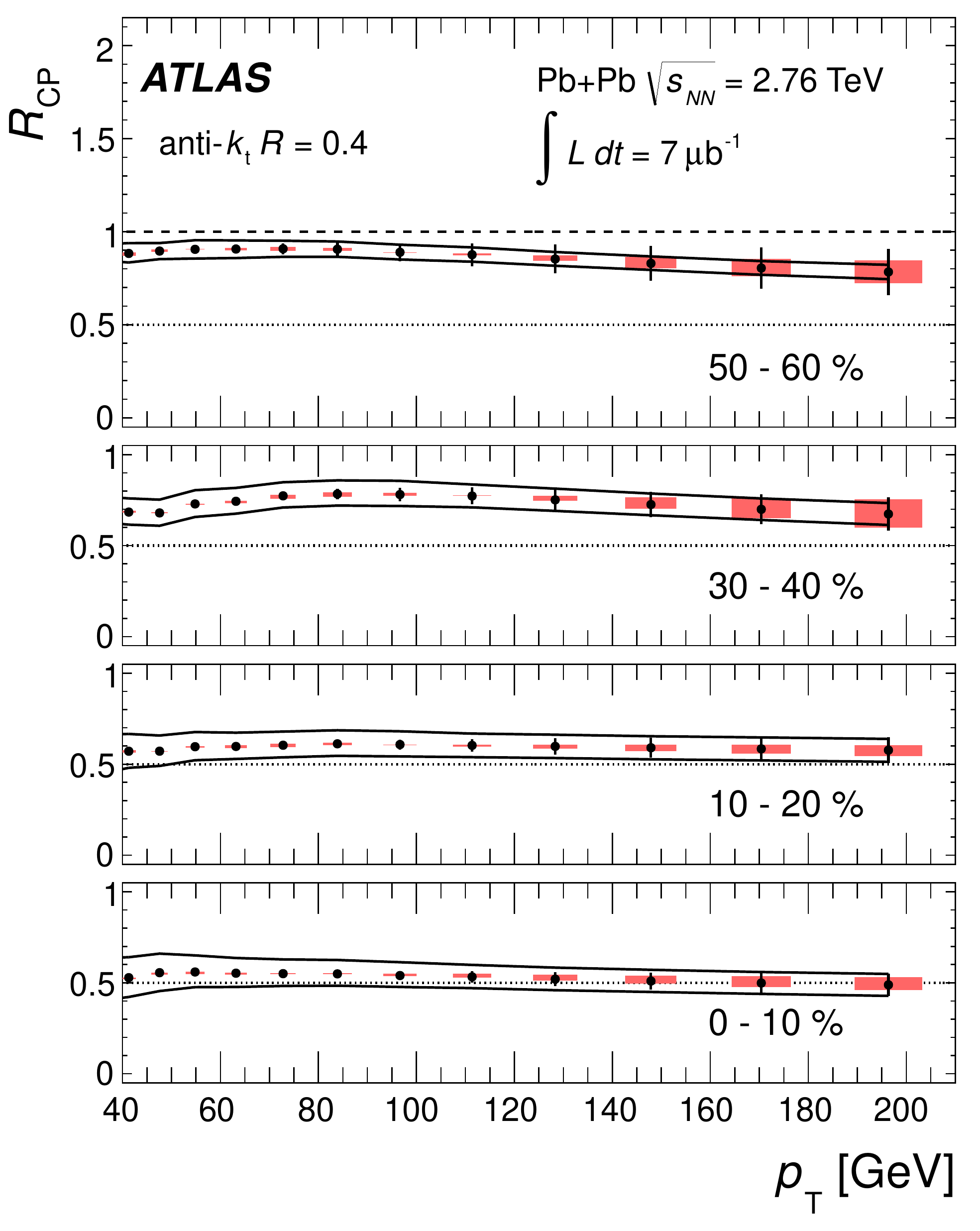}
\end{tabular}}
\caption{
\Rcp\ as a function of jet $p_{\mathrm{T}}$ for \RTwo\ (left) and \RFour\ (right) jets  
in four centrality bins. The error bars indicate
statistical errors from the unfolding. The band between the black lines indicates fully correlated uncertainties, shaded boxes partially correlated uncertainties.
}
\label{Fig3}
\end{figure}

\looseness-1 The resulting $R_{\mathrm{CP}}$ as a function of $N_{part}$ for R=0.4 jets and six jet $p_{\mathrm{T}}$ bins is shown in Fig.~\ref{Fig2}. A factor of 2 suppression in central events with respect to peripheral collisions is observed. \Rcp\ values decrease almost linearly with \Npart\ for high $p_{\mathrm{T}}$, while in low $p_{\mathrm{T}}$ ranges the \Rcp\ drops much faster with $N_{part}$ and then becomes almost constant. The jet $p_{\mathrm{T}}$ dependence of $R_{\mathrm{CP}}$ for R=0.2 and R=0.4 jets is presented in Fig.~\ref{Fig3}. The jet $R_{\mathrm{CP}}$ is observed to be similar for R=0.2 and R=0.4 and indicates only a weak variation with the jet $p_{\mathrm{T}}$.  

\begin{figure}[ht]
\centerline{
\includegraphics[width=5.9cm]{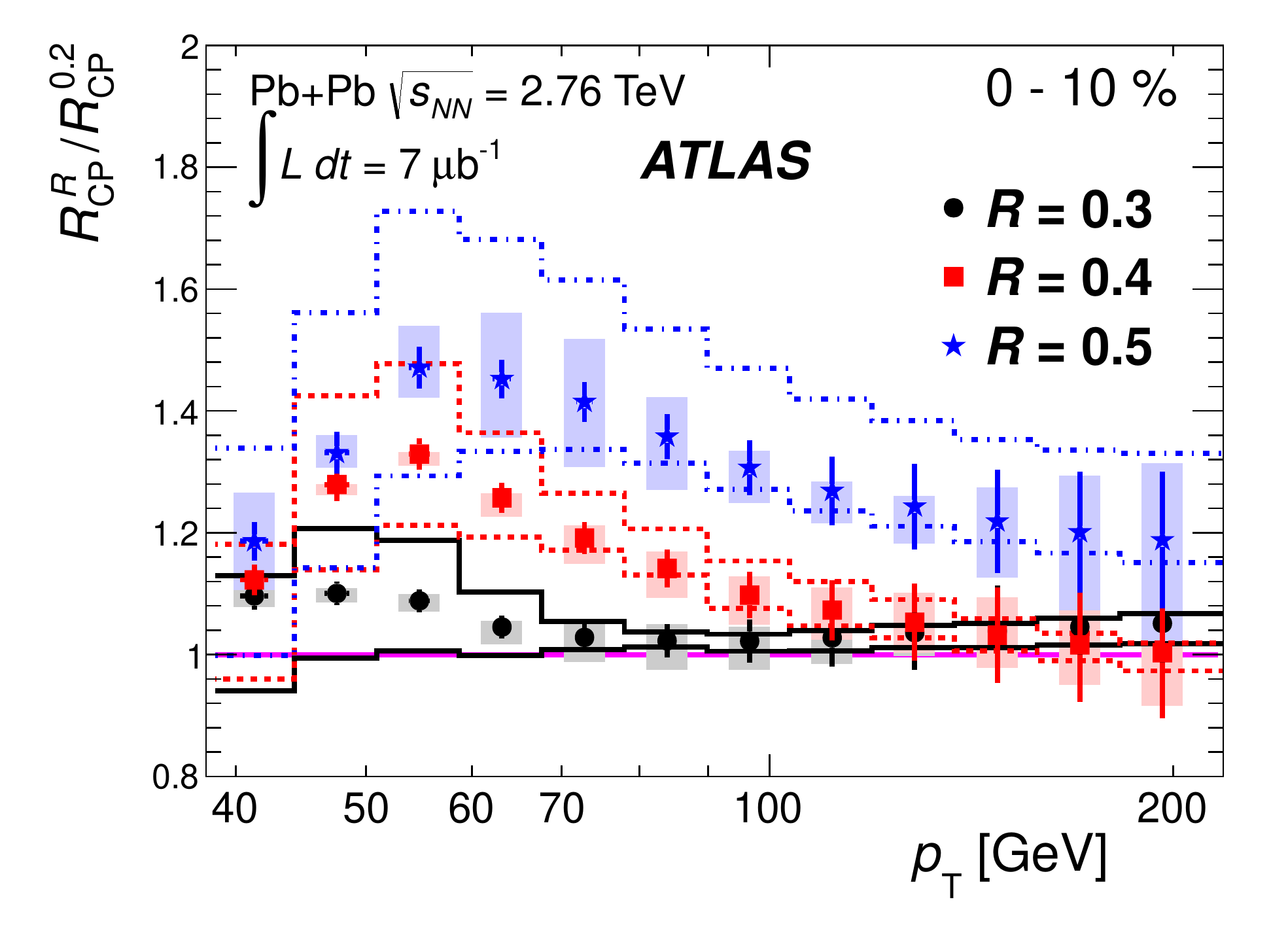}
}
\caption{
Ratios of jet \Rcp\ between R = 0.3, R = 0.4, R = 0.5 and R=0.2 jets as a function of jet $p_{\mathrm{T}}$ in the 0--10\% centrality bin. The error bars show statistical
uncertainties. Lines indicate fully correlated uncertainties, shaded boxes partially correlated uncertainties between different $p_{\mathrm{T}}$ bins.
}
\label{Fig4}
\end{figure}

\looseness-1 Predictions of radiative energy loss models suggest that the energy can be recovered by expanding the jet size~\cite{Vitev}. The study of the dependence of \Rcp\ on the jet radius is presented in Fig.~\ref{Fig4} where ratios of \Rcp\ between R=0.3, R=0.4, R=0.5 jets and R=0.2 jets as a function of $p_{\mathrm{T}}$ for central collisions are shown. A significant dependence beyond the systematics uncertainties is observed for low $p_{\mathrm{T}}$ values. This result is in a qualitative agreement with the theory predictions. More details can be found in~\cite{RCP}.

\section{Jet structure}

\looseness-1 A modification of jet structure was predicted by different theoretical models before the first Pb+Pb LHC data~\cite{Frag}. We study the jet fragmentation for R = 0.2, R = 0.4 jets in the pseudorapidity range $|\eta|<2.1$. Only charged particles with $p_{\mathrm{T}}>$ 2 GeV were used, more details can be found in~\cite{Subtraction}.

\begin{figure}[ht]
\centerline{
\begin{tabular}{ccc}
\includegraphics[width=6.4cm]{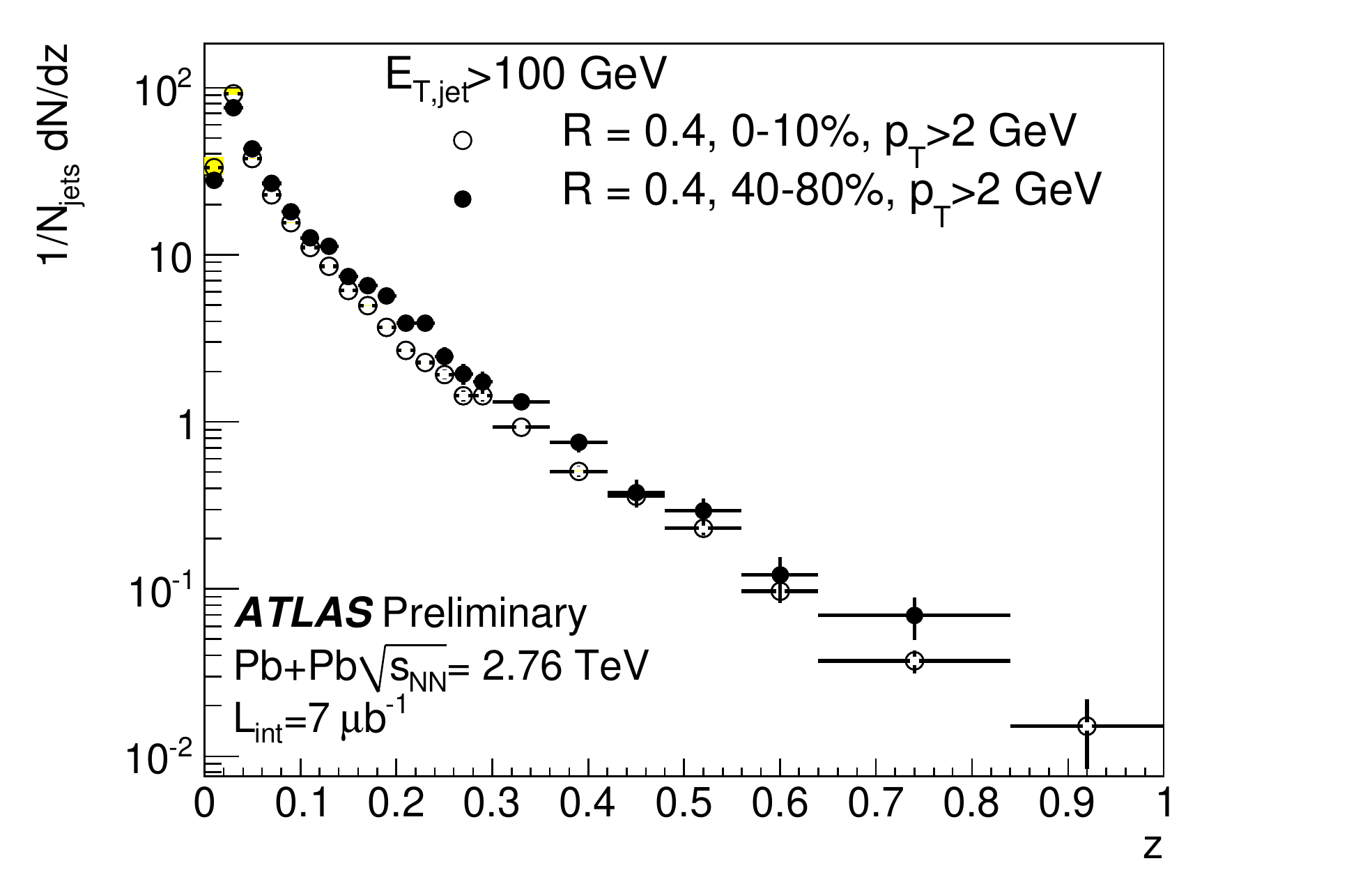}
\includegraphics[width=6.4cm]{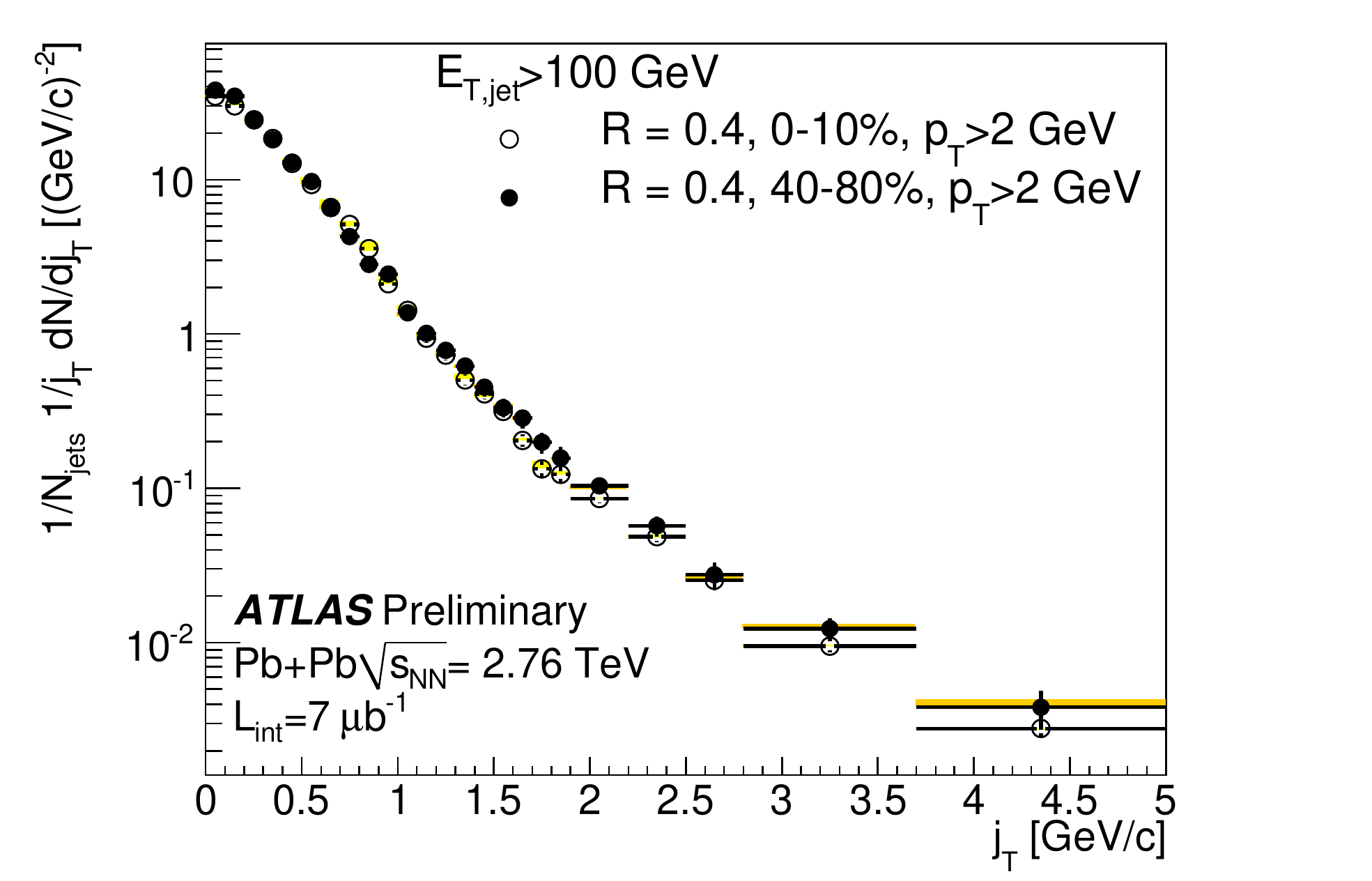}
\end{tabular}}
\caption{
Distribution of charged particles $z$ (left) and $j_{\mathrm{T}}$ (right) for \RFour\ jets with $E_{\mathrm{T}}>$100 GeV in
central (0-10\%) and peripheral (40-80\%) collisions. The yellow and orange bands show the systematic uncertainty 
from the subtraction of the UE contribution.
}
\label{Fig5}
\end{figure}

The longitudinal structure is described by the fragmentation function $D(z) = (1/N_{jet})\mathrm{d}N/\mathrm{d}z$, where\\ $z=(p_{\mathrm{T}}^{had}/E_{\mathrm{T}}^{jet})\cos\Delta R$ and $\Delta R$ is the distance between a hadron and the
jet axis in $\eta - \phi$ space. The left panel of Fig.~\ref{Fig5} shows fragmentation function for R=0.4 jets with $E_{\mathrm{T}} >$ 100 GeV in central and peripheral collisions. We observe no modification of fragmentation functions in central collisions
with respect to peripheral collisions at high $z$. A suppression at intermediate $z$ and enhancement
at very low $z$ can be seen. The transversal structure is described by $j_{\mathrm{T}}$ distribution $D(j_{\mathrm{T}}) = (1/N_{jet})(1/j_{\mathrm{T}})\mathrm{d}N/\mathrm{d}j_{\mathrm{T}}$, where $j_{\mathrm{T}}=p_{\mathrm{T}}^{had}\sin\Delta R$. The right panel of Fig.~\ref{Fig5} shows $j_{\mathrm{T}}$ distribution for R=0.4. No significant broadening of $j_{\mathrm{T}}$ distribution in central collisions is observed.

\section{Summary}
\looseness-1 We have presented study of the jet suppression and measurement of two fragmentation variables. Results were obtained by the ATLAS detector using 2010 Pb+Pb data at $\sqrt{s_{NN}} = 2.76$~TeV. A suppression by a factor of 2 is observed in the jet yield in central with respect to peripheral collisions. The dependence of the jet suppression on jet $p_{\mathrm{T}}$ is very weak. A significant dependence of the suppression on the jet size is observed for low jet $p_{\mathrm{T}}$ values. We do observe a significant modification at low and intermediate $z$. No significant modification is seen at high $z$.

\bibliographystyle{elsarticle-num}
\bibliography{lit}

\end{document}